# Visualizing the Invisible Hand of Markets:
# Simulating complex dynamic economic interactions

Klaus Jaffé

**Universidad Simón Bolívar, Caracas, Venezuela**

kjaffe@usb.ve

**Abstract:** In complex systems, many different parts interact in non-obvious ways. Traditional research focuses on a few or a single aspect of the problem so as to analyze it with the tools available. To get a better insight of phenomena that emerge from complex interactions, we need instruments that can analyze simultaneously complex interactions between many parts. Here, a simulator modeling different types of economies, is used to visualize complex quantitative aspects that affect economic dynamics. The main conclusions are: 1- Relatively simple economic settings produce complex non-linear dynamics and therefore linear regressions are often unsuitable to capture complex economic dynamics; 2- Flexible pricing of goods by individual agents according to their micro-environment increases the health and wealth of the society, but asymmetries in price sensitivity between buyers and sellers increase price inflation; 3- Prices for goods conferring risky long term benefits are not tracked efficiently by simple market forces. 4- Division of labor creates synergies that improve enormously the health and wealth of the society by increasing the efficiency of economic activity. 5- Stochastic modeling improves our understanding of real economies, and didactic games based on them might help policy makers and non specialists in grasping the complex dynamics underlying even simple economic settings.

# INTRODUCTION

Adam Smith (1776) and Friedrich Hayek (1961), among many others, assigned extraordinary importance to hidden properties of markets. They assumed the existence of poorly understood complex market mechanisms upon which modern economies are build. These mechanisms and the phenomena they produce, are not easily grasped analytically because of their extraordinary complexity. Mathematics, the language we use to ask questions to nature, has expanded the limits of economic analytical power since Hayek wrote about the "Economic Calculus" and Smith about "The Invisible Hand". We now are capable of viewing economies with more powerful tools that allow for a more rigorous analysis of complex features formerly regarded as off limits to rational reductionist methods by economists. These tools allow producing multidimensional radiographies of an economy. They include "Cellular Automata" (Axelrod 1984), "Active Walks" (Lam 2005, 2006) and computer simulations. Specifically, ABM or Agent Based Modeling, allow simulating very complex phenomena in economics (Prietula & Carley 1994, Kochugovindan & Vriend 1998, Magliocca et al 2014, for example). ABM simulations can easily be made so complex as to be beyond our understanding, becoming unpredictable. Thus, in order to remain useful, we have to limit the complexity of ABM so as not to trespass this edge of chaos. The mathematically relevant challenge is to find the simplest representation of reality that able to capture the problem to be studied.

Sociodynamica is an agent based simulation model, purposely build for analyzing economic micro-dynamics. It is based on classical monetary economic principles (Gurley & Shaw 1960, Wray 1996) and builds virtual societies of agents that collect and trade two types of heterogeneously distributed resources. It can be made as complex as wished, but in order to comply with the maximum entropy principle (Jaynes 1957) we want to keep the problem simple, and advance our knowledge in small steps. The aim here is to build the simplest mathematical construct that still reflects aspects of the economic dynamics that are fundamental to modern complex economies. Specifically we want to reproduce phenomena, such as the emergence of the "Invisible Hand", as coined by Adam Smith. Sociodynamica is specifically aimed to visualize the process of emergence of new macro phenomena from the microscopic social interaction of agents, and has been in continuous development, starting from Biodynamica, for the last 20 years. A large number of alternative agent based simulation models for economic problems are available (see Tesfatsion 2014). For example, "Sugarscape", an agent based model developed independently by Axelrod (1997), or the model developed by Axtell (2005), are very similar to Sociodynamica. Eventually, repeating the simulations presented here with other agent based models, and devising empirical studies based on these simulations, should add confidence to the results presented and advances our understanding of complex markets.

**METHOD**

Sociodynamica is a freely available agent based simulation model written in Visual Basic in which diverse agents roam a virtual landscape looking for resources. Agents farm and mine for foods and minerals and trade their surplus according to different economic settings. Sociodynamica assumes that utility functions in economics are equivalent to fitness functions in biology (Kenrick et al. 2009), simulating the survival of agents in situations with or without heredity and with or without movement of agents. Sociodynamica, has been used previously to study the effect of altruism and altruistic punishment on aggregate wealth accumulation in artificial societies (Jaffe 2002a, 2004a, 2008, Jaffe & Zaballa 2009, 2010) and in studying the dynamics of complex markets (Jaffe 2002b, 2004b).

The present model simulated a virtual society of agents that farmed and mined food and minerals respectively, and traded their surplus according to different economic settings. The agents inhabited a continuous flat two-dimensional toroidal world (see Figure 2) that was supplied with patches of agricultural land ("sugar" or "food") and separate non-overlapping patches of mines ("spices" or minerals). Immobile agents were randomly dispersed over this fine-grained virtual landscape. Each time step, agents that happened to be located over one of the resources, acquired a unit of the corresponding resource, accumulating wealth, either as sugar or food ($G_1$) and/or as spices or minerals ($G_2$). Agents spend a fixed amount of each resource in order to survive, consuming each of them at a basal constant rate (default value was 0.1 units of the corresponding resource). Both resources were consumed and metabolized similarly, but food was 3 times more abundant than minerals (the size of the patch for minerals was normally set to 100 x 100 pixels and for food 300 x 300 pixels). Each patch remained in the same place during each simulation run and the resources inside them were replenished continuously. Agents perished when they exhausted any of the two resources. The population of agents was maintained constant by introducing after each time step new agents with randomly assigned initial parameters. Initial parameters were: the type of agent, the random spatial position and the initial amount of money used to start trading resources (the default initial value was set to 10 units of money). The amount of money each agent possessed varied according to its trade balances. Agents gained money when selling food and/or minerals and spend money when buying them.

Agents traded the resources they possessed with other agents. They sold the resource they possessed in more abundance and bought the one they were short of. In simulations without "division of labor", the trade could take place between any agents in a population of omnipotent agents. In simulations with division of labor, agents traded according to their specialization. Here, agents were subdivided into three categories: Farmers which specialized in collecting only Food (resource 1); Miners which extracted only Minerals (resource 2); and Traders specialized in trading minerals for food when encountering a Farmer, and food for minerals when encountering a Miner. Farmers traded only with Miners and Traders; Miners traded only with Farmers and Traders. Traders could interchange resources with all types of agents. Trades were allowed only between agents spaced at a distance not larger than the "Contact Horizon" of the trading agent. Each time step, buyers searched for potential sellers of the required good by contacting randomly up to 10 agents in the area defined by this Contact Horizon. If the agent found another agent with the wanted good willing to sell at or below the price defined by the buyer, a trade was executed using the price of the seller. Trades were limited to the amount of money available to the buyer and the amount of goods possessed by the seller, unless credit was simulated. Variation of the contact horizon allowed us to simulate different levels of globalization or integration of economic agents. Sellers not able to sell their product during one such trading session decreased their price for the resource; and buyers unable to get what they wanted at the preferred price increased the price they were willing to pay in the next round of trades.

Sociodynamica can simulate biological and/or cultural evolution. Both have similar dynamics (Jaffe & Cipriani 2007). Here we simulated cultural evolution in economic markets. This evolutionary process weeded agents with unfit combination of parameters revealing the optimal set for each economic scenario. The time immobile agents survived was reflected in the average age of agents or "Age" and was used as an estimate of the "Health" of the economy. The "Wealth" of the economy was assessed by the total resources (Food and Minerals) accumulated by surviving agents.

*Type of resources (Res)*

Two different types of interactions between agents and resources were modeled. To make results comparable to popular alternatives to Sociodynamica, we simulated the setting of "Sugar and Spices" (**SS**) by Axelrod (1984). In simulations of SS, both resources were consumed and metabolized similarly. When simulating Food and Security (**FS**), Security was provided by Resource 2, and the wealth in resource 2 ($G_2$) was inversely related to the probability of sudden death for each agent. That is, Resource 2 improved the odds of surviving external "catastrophes" that killed agents at random from time to time. A large amounts of $G_2$ protected the agents against these catastrophes by reducing the probability of being affected by them so that agents survived if $G_2 - Rnd(0-1) * D > 0$

This algorithm achieved that the greater the wealth of accumulated minerals of the agent, the lower their probability of being struck by a catastrophe, at any level of danger (D).

Money served to trade resources. Each agent was endowed wit a initial amount of money (10 units unless stated otherwise) which was used to pay for sales and to provide credit.

*Simulations*

Different aspects of an economy were modeled including:

1. Credit for agents lacking money (TE1)
2. Biased influence of prices biased towards sellers (TE2)
3. Biased influence of prices biased towards buyers (TE3)
4. Different relative importance between the two resources (abundance regulated through patch size or consumption regulated by metabolism)
5. Different geographical (spatial) distribution of resources (Topology of the distribution)
6. Different balances between income and use of resources (balancing productivity of resource with its consumption)
7. Different type of resources: Sugar and Spices where the two resources behaved similarly, or Food and Security the effect of one resource on fitness was constant and the other (security) was random and variable
8. Prices an agent was wiling to pay for a given resource increased after each failed attempt to buy, and decreased after each failed attempt to sell the respective resource. In some simulations specifically mentioned below, agents, in addition increased the price to sell after a successful transaction, or decreased the price to buy after a successful shopping.

9. Combinations of the above

Economic equilibrium of a given modeled system was assessed using the following variables:

Balance between income (I) of resources (r) and their consumption (C). For survival, agents were required to conform to

$$Ir > Cr$$

Income can be either by direct gathering (G) or by trade (T)

$$Ir = Gr + Tr$$

Here each agent has to balance two resources in order to survive. I simulated a utility functions (U) so that U had to remain positive for the agents survival and:

$$Ur = Gr + Tr - Cr$$

The amount of traded resources can increase or decrease, according to the balance of resourced bought (B) and sold (S)

$$Tr = Br - Sr$$

The amount of resources bought and sold depends on the availability of money (M) and the price (P) paid for the resource by the agent (a)

$$Br = Ma/Pr_a \quad \text{and} \quad Sr = Ma/Pr_a$$

The amount of money of agent (a) depends on the amount spend buying (Mbr) and the amount gained selling (Msr) resources

$$M_a = M_{a0} + Msr - Mbr$$

where $M_{a0}$ was the initial amount of money supplied to each new agent j

The computer simulations consisted in numerically calculating $Ur_a$ every time step for the population of agents (a) so that the total accumulated of wealth for each resource (Wr)

$$Wr = \Sigma_a Ur$$

Agents with $U_1 <= 0$ or $U_2 <= 0$ were eliminated and substituted by new ones with default properties, as an analogy of broken companies that are replaced by new start-ups.

Results are presented with only four variables that reflect the state of the simulated economies:
  1. **TAR**: Total amount of each resource accumulated by the population. TAR = $W_1 + W_2$

2. **Age**: Average age of agents measured in Time steps.
3. **FPr**: Average price for resource 1, called Food, equivalent to either Sugar or Food.
4. **MPr** Average price of resource 2, called Mineral, equivalent to either Spices or Security.

The variables or features of the simulations explored here are summarized in Table 1.

**Table 1:** Features explored in the simulations

| SS | Resources behaved as in Sugar and Spices |
|---|---|
| FS | Resources were Food and Security, were Minerals protected against random culling |
| Type of Agents | Omnipotent: All agents collected and traded both type of resources<br>Division of labor: Agents were Farmers or Miners, specialized in collecting and selling food or minerals, or Traders specialized in trading both resources |
| FMar | Price adjustment in monetary units. 0 is equivalent to fixed prices and 0.5 when prices were flexible. |
| FPro | Flexible productivity so that increased prices increased production of resource |
| TE | Type of economic feature simulated.<br>TE0 = Control simulations with default parameters<br>TE1 = Credit<br>TE2 = Agents finishing a successful sale, increase price by one unit<br>TE3 = Successful buyers decrease their future asking price by one unit<br>TE4 = 2 + 3 |
| Danger D | Each time step of 10 % of the population was killed, unless another value is given. Relative wealth in minerals protected agents. The algorithm used to trigger culling was: $100 * Rnd / U_2 / W_2 < 1000 * Rnd / D$;<br>D represented the perceptual value assigned for culling. |
| Credit | When agents buying from a trader or from an omnipotent agent had less money than required to finish the transaction, they asked for a loan. The trader gave half of the amount of the resource requested for trade to the buyer and the buyer increased its depth according to the price of the resource given by the seller. After this trade, each time the buyer met a trader and if it had the money, it paid two monetary units to this trader, which could be different to the former one, until the depth was canceled. |

The detailed program in Visual Basic is available in the help feature of the program. Simulations can be run with parameters choose at will by downloading Sociodynamica at [http://atta.labb.usb.ve/Klaus/Programas.htm]. Here I present the results of only 16 types of simulations, each type run at lest 200 times for 200 time steps, as a representative sample of the effects of the various parameters studied. Default values, unless explicitly stated otherwise were Transaction

horizon = 200, Initial prices for both resources = 3 units, Reserve units not traded = 1 Patch size for resource 1 = 300, Patch size for resource 2 = 100, Resources metabolized per time step = 0.1, Resources collected per time step if placed over the resource = 2. Number of agents = 500, maximum number of trades per time step = 10.

More details are given in the appendix. Videos of simulations are available at: http://atta.labb.usb.ve/Klaus/EC/ECVideos.html

**RESULTS**

The simulator allowed the design of experiments where several different variables interacted with each other simultaneously, so as to study complex interactions between several factors affecting the economy at the same time. The results of over two hundred thousand simulations exploring the effect of dozens of interrelated variables cannot be presented here in detail. For a fuller exploration of the simulations, the reader is advised to download the simulation program and design experiments on his own. Only a selected summary of lessons learned from Sociodynamica's simulations is present. These can be grouped into results that confirm known economic dynamics, validating the model (results 1-3); results that are relevant for understanding economic micro-dynamics (results 4-5); and novel serendipitous or unexpected findings (results 6-7):

**1. Wealth distribution among agents mirrors that of empirical data**

The wealth distribution among agents showed to be a negative exponential function with fat tails (Figure 1). This form of the curve is indistinguishable from those found empirically by economists and physicists (see Levy & Solomon 1997, Dragulescu et al 2000, Gonzalez-Estevez et al 2008, Lopez-Ruiz et al 2012, for example). The exact form of the curve varied somewhat showing slight differences when obtained in simulations with differed parameters. But all distributions obtained showed the same basic power distribution with long tails. This distribution differs from a normal distribution, which is assumed by default by most classical economist when analyzing economic data. Averages and parametric statistics are not very meaningful in analyzing data with these distributions. Non-parametric statistics is more adequate for handling these kinds of data.

**Figure 1.** Example of frequency distribution among 2000 agents in respect to different amounts of goods possessed by each agent in a selected simulation.

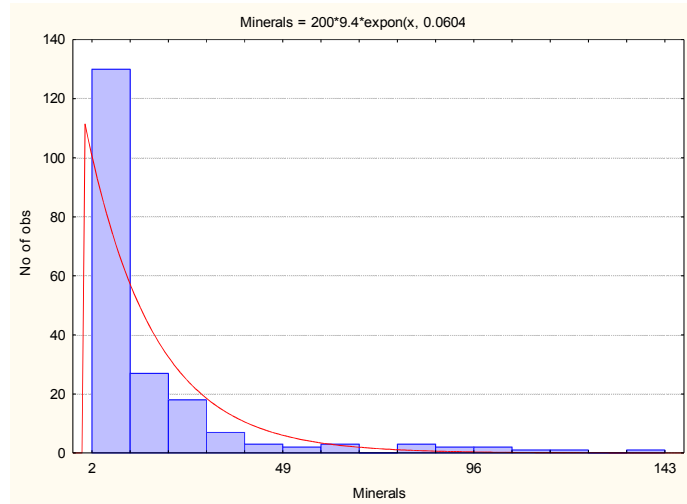

## 2. Heterogeneous distribution of resources affects price dynamics

Prices, when allowed to be adjusted by individual agents, were heterogeneously distributed among the population of agents, and sellers of a resource had higher price expectation than buyers. That is, sellers of sugar were more likely to expect sugar to be dearer than spices (agents with a red aura in Figure 2) and sellers of spices were more likely to expect spices to be dearer than sugar (blue aura around agent in Figure 2). This effect was due to the fact that prices increased with scarcity. Prices for the two goods were similar (simulations achieved for both prices of 1.2 units after 200 time steps; a t-test revealed no difference between both prices at p<90) when land with spices had the same area than land covered with sugar. When land covered with sugar had 3 times the area than land covered with spices, simulations showed that prices for the scarce resource, spices, were higher (1.7 units after 200 time steps) compared to prices of the more abundant one, sugar (1.2 units). The difference was statistically highly significant (p<0.0001 using a t-test comparing both prices)

**Figure 2**: Examples of total resource landscape and amplification of a section showing agents. First row; examples of simulations EC01, EC02; second row, EC09, EC15.

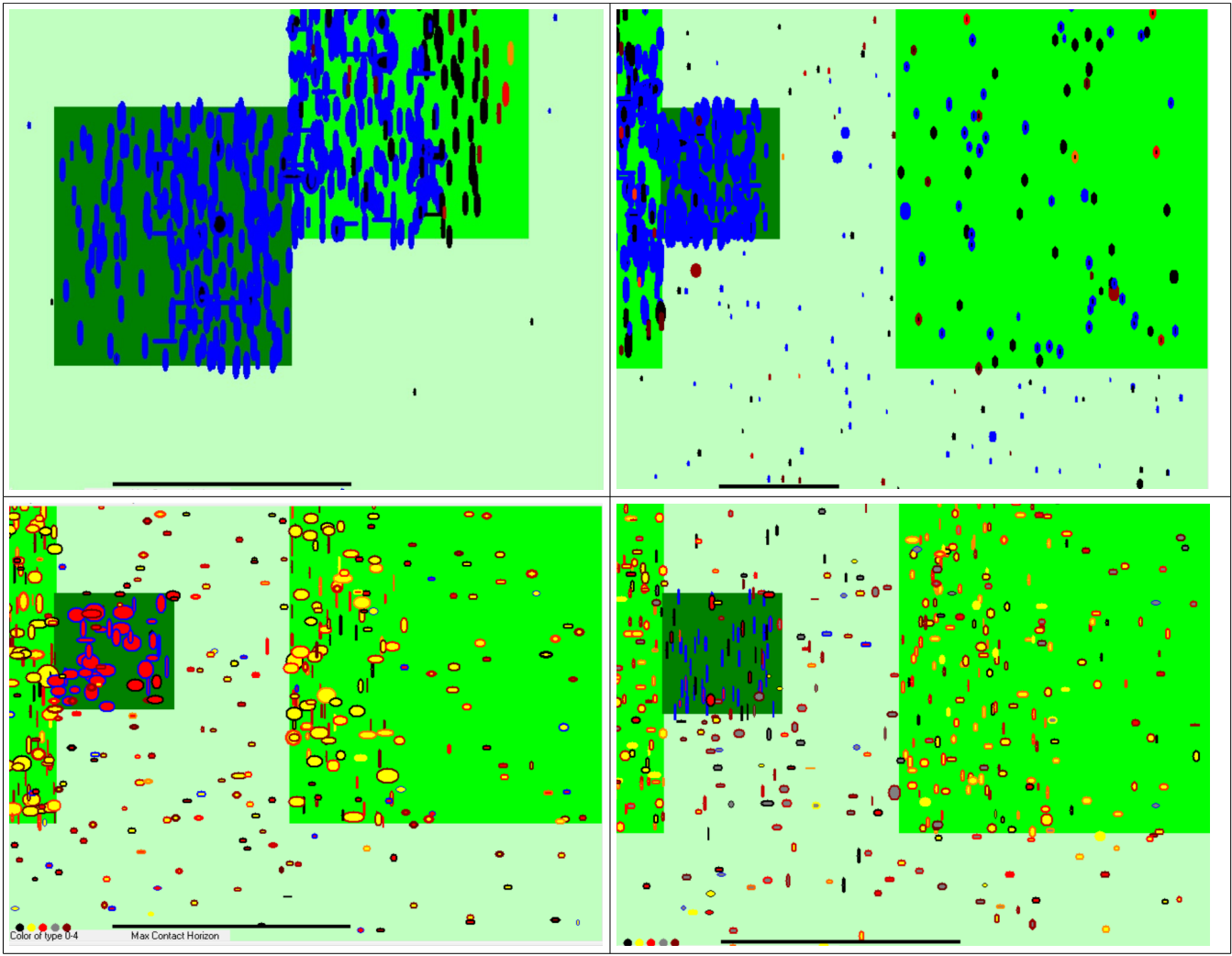

- Bright green field is covered with "Food"; darker green field is covered with "Minerals"; the lightest green is devoid of resources.
- The size of the bubble is proportional to the wealth of the agent so that the thickness is determined by ($Ur_1 + Ur_2$) and the height by the amount of money the agent possesses.
- The thickness of the border is proportional to the perceived cost of living calculated as $FPr + MPr$
- The color of the border is more reddish or even yellow the higher the ratio $FPr/MPr$. That is, agents with red and yellow pay more for minerals, whereas those with blue or black pay less for minerals compared to what they are willing to pay for food. For simulation EC001 and EC002 using omnipotent agents with no division of labor, this same color was applied to the whole body. For the rest of the simulations, the color of the body was proportional the color of the body of sphere described the type of agent so that yellow were Farmers, red were Miners and grey were Traders.

## 3. Spot prices are much more volatile than average prices

The model allowed tracking simultaneously spot prices and the average acceptable prices of all agents. This is normally not possible in real economies. The spot price was calculated from the last two transactions after every time step; whereas averages were calculated by taking into account all transactions during a time step. Average prices were an estimate of the price agents were willing to pay in a commercial transaction. Spot prices and average prices differed significantly (Figure 3). The specific distribution of price differences between food and minerals depended on the economic topology of the system, but spot prices were always more volatile. The long-term average for spot prices was similar to the average prices for each resource in the population. However, when selecting randomly a given spot price, chances were that spot prices differed largely from the average prices.

**Figure 3:** Spot and PR are the ratio of Food Price (FPr) / Mineral price (MPr) for the last two transactions (Spot) or for the average of the whole population of agents

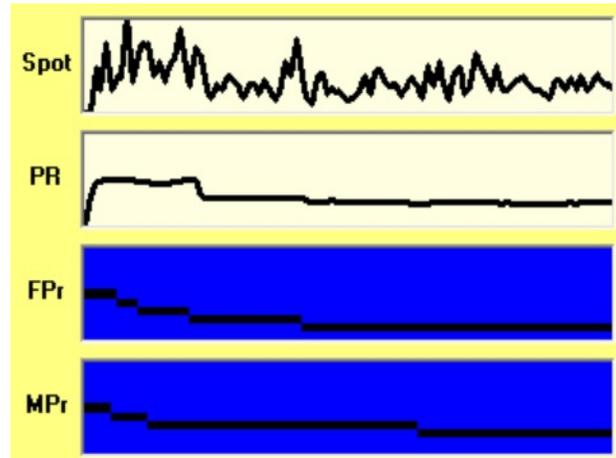

**4. Elements affecting the wealth and health of an economy**

Table 2 presents the effect of the different features simulated on the health and wealth of the virtual economy. Health was assessed using "Age", which reflects the quality of life for the average agent, measured through the average age of the agents. Age was correlated to a measure of wealth of the economy called "TAR" which represents the total amount of resources (food and minerals) accumulated. The most important features revealed by the results in Table 2 were:

**Table 2:** Wealth (TAR in thousands of units) and Health (Age in time steps of the average agent) in different economies with 500 agents after 200 time steps. Res indicates the type of resources simulated, SS for sugar and spices and FS for food

and security; FMar indicates if prices could vary; FPro indicate if productivity was flexibly regulated by prices. Omnipotent agents simulate no division of labor. Sim refers to the specific simulation.

| Omnipotent agent | | | TAR | | | | | Age | | | | | Sim |
|---|---|---|---|---|---|---|---|---|---|---|---|---|---|
| Res | FMar | FPro | TE0 | TE1 | TE2 | TE3 | TE4 | TE0 | TE1 | TE2 | TE3 | TE4 | |
| SS | n | n | 3 | 8 | | | | 2 | 10 | | | | EC04 |
| SS | y | n | 4 | 8 | 3 | 6 | 4 | 4 | 9 | 2 | 6 | 3 | EC05 |
| SS | y | y | 5 | 4 | 4 | 3 | 5 | 11 | 8 | 4 | 7 | 9 | EC13 |
| FS | n | n | 5 | 4 | | | | 7 | 5 | | | | EC06 |
| FS | y | n | 5 | 4 | 5 | 5 | 5 | 7 | 5 | 6 | 6 | 7 | EC07 |
| FS | y | y | 3 | 3 | 6 | 3 | 4 | 7 | 5 | 6 | 5 | 7 | EC12 |
| Division of labor | | | | | | | | | | | | | |
| SS | n | n | 58 | 63 | | | | 73 | 78 | | | | EC08 |
| SS | y | n | 49 | 44 | 51 | 48 | 51 | 78 | 93 | 71 | 76 | 73 | EC09 |
| SS | y | y | 11 | 9 | 44 | 3 | 25 | 89 | 96 | 70 | 64 | 74 | EC15 |
| FS | n | n | 8 | 9 | | | | 17 | 17 | | | | EC10 |
| FS | y | n | 4 | 4 | 6 | 3 | 4 | 15 | 17 | 16 | 14 | 15 | EC11 |
| FS | y | y | 2 | 2 | 7 | 1 | 3 | 14 | 15 | 18 | 14 | 13 | EC14 |

TE0 = Control; TE1 = Credit; TE2 = Agents finishing a successful sale, increase price by one unit; TE3 = Successful buyers decrease their future asking price by one unit; TE4 = TE2 + TE3; TE5 = TE4 + TE1. F Mar = y then prices changed 0.5 units each time a trade failed. Simulations of TE2 to TE4 are not possible when prices are fixed (FMar = n).

*4.1 Synergy is created by division of labor*

When we compare the results of simulation where agents divided their labor with those where agents were omnipotent, we observe that division of labor increased price differentials, longevity and wealth. Specifically, there was a very substantial difference in TAR and Age. The division of labor created synergies in the economic dynamic, fomenting useful trades, i.e. between collectors of food with collectors of minerals, and avoiding useless trades, i.e. between collectors of the same resource type. The difference was higher among simulations using SS settings than among those using FS settings.

*4.2 Type of resources simulated*

The type of resources simulated affected the outcome of the simulation greatly. Simulating sugar and spice (SS, data in blue rows) produced different outcomes compared to simulations with food and security (FS). Wealth levels achieved by agents and their longevity were much larger with SS than with FS. However, the decrease in longevity caused by FS was worse without division of labor. Simulations with omnipotent agents, i.e without division of labor, produced much lower levels of wealth and longevity.

*4.3 Credit*

Simulating credit increased the accessibility of resources significantly when omnipotent agents were simulated (EC04, EC05). In contrast, the effect of credit was very small when division of labor was modeled in the simulation (EC08, EC09). That is, the positive effect of credit was less evidenced when the positive effect of synergies triggered by division of labor was present. When prices were flexible, the positive effect of credit was reduced or eliminated (for example EC5 vs EC13), reflecting the fact that improved commercial transactions are better achieved by flexible pricing then by increasing money supply through indiscriminate credit as modeled here.

*4.4 Correlations between Wealth and Health*

In general, wealth and health reflected similar aspects of the economy. The correlation between both variables in the data shown in Table 2 is high ($r = 0.82$), though both measures were not identical. In some cases "Age" and "TAR" were not correlated. An interesting effect can be observed in economies with division of labor, where a fuller economic development was achieved: Focusing on TE0, we observe that flexible productivity decreased TAR but increased Age (EC09 vs EC15). In contrast, flexible prices (FMar = yes) increased both health and wealth of the societies (compare EC10 vs EC15).

**5. Elements affecting prices**

The dynamics of prices differed from that of TAR and Age. In general, prices behaved as expected (Table 3). Some interesting features about prices revealed by these simulations were:

**Table 3:** Prices for Food (FPr) and Minerals (MPr) in simulations in Table 1. DL stands for division of labor

| Prices | | | FPr | | | | | MPr | | | | | |
|---|---|---|---|---|---|---|---|---|---|---|---|---|---|
| Res | DL | FPro | TE0 | TE1 | TE2 | TE3 | TE4 | TE0 | TE1 | TE2 | TE3 | TE4 | Sim |
| SS | n | n | 0.9 | 0.9 | 1.7 | 0.6 | 1.0 | 1.1 | 1.1 | 2.0 | 0.7 | 1.2 | **EC05** |
| SS | n | y | 1.0 | 0.9 | 1.6 | 0.5 | 1.0 | 2.2 | 1.0 | 3.4 | 0.5 | 2.0 | **EC13** |
| FS | n | n | 1.0 | 0.9 | 2.9 | 0.6 | 1.2 | 1.6 | 1.2 | 2.8 | 0.6 | 1.1 | **EC07** |
| FS | n | y | 0.9 | 0.8 | 2.3 | 0.5 | 1.2 | 1.4 | 1.1 | 2.2 | 0.6 | 1.2 | **EC12** |
| SS | y | n | 1.0 | 0.9 | 1.8 | 0.5 | 0.9 | 3.4 | 3.0 | 2.9 | 0.6 | 1.0 | **EC09** |
| SS | y | y | 0.8 | 0.8 | 1.8 | 0.5 | 0.9 | 3.2 | 3.1 | 2.8 | 0.5 | 1.0 | **EC15** |
| FS | y | n | 0.9 | 0.9 | 2.0 | 0.5 | 1.0 | 4.3 | 4.8 | 3.1 | 0.6 | 1.0 | **EC11** |
| FS | y | y | 0.8 | 0.8 | 2.5 | 0.5 | 1.1 | 3.7 | 3.9 | 3.9 | 0.6 | 1.0 | **EC14** |

*5.1- Heterogeneous prices reflect the efficiency of an economy*

Division of labor increases the price difference between Food and Mineral. Mineral prices were always higher than Food prices. This reflects the fact that the smooth working of an economic system helps produce better pricing, which in turn favors the economy. This explains why prices of both resources increased with FS simulations compared with SS simulations, as with FS type of resources the economic dynamic was more chaotic.

*5.2- Credit did not affect prices*

Although credit affected the health and wealth of society (Table 2), it did not affect prices (Table 3). Credit increased the pool of total resources accumulated by agents and their average age, but had little effect on prices (compare values under TE0 vs TE1).

*5.3- Psychology of traders affects prices*

The disposition to lower the price at which an agent was willing to buy in the next round of trades had a larger effect on the dynamics than price increases implemented by successful seller when buyers outnumbered sellers. In Table 3 prices in the column for TE2 were always higher compared to TE3, independently of the balance between supply and demand. Asymmetries in price sensitivity between buyers and sellers will produce above average price inflation. The results showed that stingy sellers increased prices and less generous buyers decreased them.

**6. The degree of globalization, i.e., the Contact Horizon of agents affects prices depending on the topology of the resource distribution.**

In this simple model, the topology of the resource distribution had an effect on prices (Figure 4). The results showed that globalization reduced prices if the increased contact horizon did not introduced additional topological barriers. Price levels for the scarce resource (minerals), were higher when the commercial horizon of the agent was smaller when barriers to trade were absent, i.e. agents could easily bridge the gap between patches with food and patches with minerals. Overall economic wealth and longevity of agents increased with increased contact horizon (Figure 5). The distribution of data in Figures 4 and 5 show that linear regressions do not capture the important elements of this dynamics.

Increasing the contact horizon of agents increased the odds of an agent to complete a successful trade. Thus, synchrony in economic transactions increased with increased contact radius. This fact explains the results obtained.

**Figure 4**: Examples of the dynamics of price for goods (circles: Sugar or Food; squares Spices or Minerals) reached after 200 time steps with different maximum Contact Horizons for agents in simulations of sugar and spices.

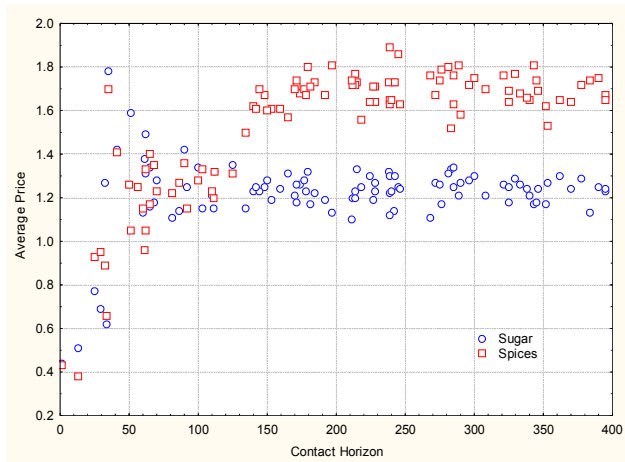

**Figure 5**: Total production of goods (Food + Minerals) and mean age achieved by economic agents always increased with increasing maximum contact horizons (EC7)

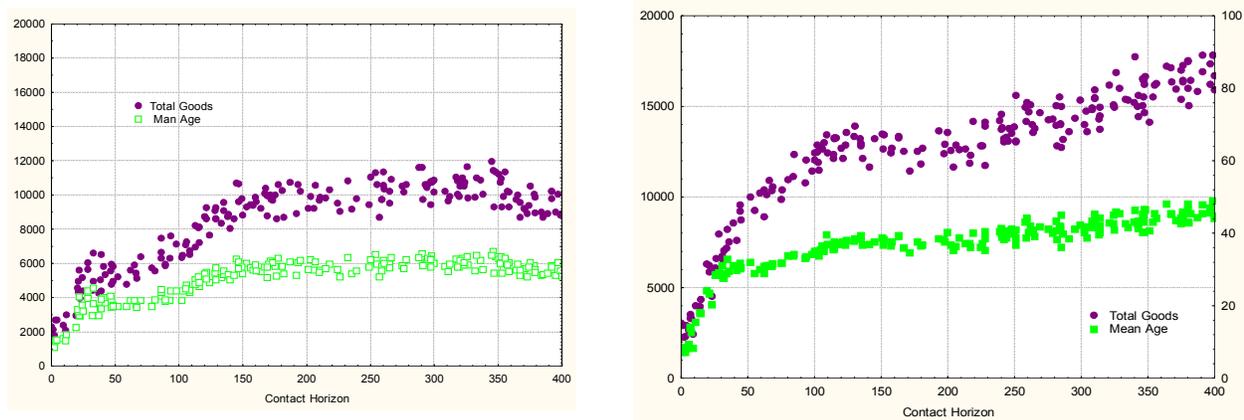

## 7. The "Invisible Hand" proposed by Adam Smith can be made visible, allowing the "Economic Calculus" supposed to be impossible by Friedrich Hayek.

Among the various factors studied, two were most relevant regarding the objective of visualizing emergent phenomena in the market. These factors were flexible pricing and division of labor. Each of these factors revealed non-intuitive features in the economic dynamics.

*7.1- Price flexibility protects the individual, and more so when credit is available*

Table 4 focuses on the effect of free markets (agents could adjust prices: Free Mar), as compared to markets with fixed prices for both resources. (no Free Mar). The results are unexpected. Free pricing in markets increased the health of a population (Age) but decreases its overall wealth (TAR). This is one of the rare situations when TAR and Age were not correlated. The presence of credit (TE1) increased this contrasting difference. That is, price flexibility favored long term survival of individuals, reflected in an increased average longevity, but reduced the overall aggregate wealth. This effect can be explained by the fact that with flexible prices, individuals can adjust their accessibility to each of the two resources, maximizing their wealth depending on the environment they were located: Miners paid more for Food and Farmers more for Minerals, maximizing their utility. With fixed prices, optimal trading habits could not be adapted to the microenvironment. But flexible prices made some resources inaccessible to some agents, reducing the total amount of accumulated resources.

**Table 4:** Simulations with fixed prices (EC08) compared to simulations with freely changing prices (EC09). Both simulations had division of labor with 3 agents and SS type of resources.

|          | TAR |     | Age |     |      |
|----------|-----|-----|-----|-----|------|
| **FMar** | TE0 | TE1 | TE0 | TE1 | Sim  |
| n        | 58  | 63  | 73  | 78  | **EC08** |
| y        | 49  | 44  | 78  | 93  | **EC09** |

*7.2. Division of labor creates economic synergies that are difficult to predict*

It could be argued that omnipotent agents are better prepared to be successful in any given economic environment as they can adapt to a wider range of circumstances, thanks to the fact that they can act as miners, farmers and traders. This however, did not to turn out to be the case. The simulation of division of labor, compared with simulations with omnipotent agents, increased the health and wealth of the population enormously (Table 5). This effect was the strongest among all the contrasting combination of possibilities explored. Its origin derives from the fact that in simulations with division of labor, agents avoid selling resources they do not collect. That is, farmers sold only food and miners sold only minerals. Traders sold whatever resource they had more abundant. The presence of traders helped populate spaces not occupied by any of the resources, as was also the case when omnipotent agents were simulated. This, though, seemed to add little to the overall wealth of the populations.

**Table 5:** The effect of division of labor on the wealth and health of populations exploiting SS resources.

| Age | Div Labor | TE0 |
|---|---|---|
| EC05 | n | 4 |
| EC09 | y | 78 |
| **TAR** | | |
| EC05 | n | 4 |
| EC09 | y | 49 |

This result shows that even in this simple example, optimal economic strategies differ for different agents. That is, Farmers do best if they only sell to Miners, and miners do best if they only sell to Farmers. Intermediate strategies failed to reap maximum wealth. This divergence of optimal strategies explains the success of division of labor, as in areas with minerals, only Miners survived, and in areas with Food, only Farmers survived after a few time steps, producing a division of labor adjusted to the heterogeneities of the heterogeneous distribution of resources.

**DISCUSSION**

The simulations presented here show that the simulator is a powerful analytical tool, producing a rich collection of outcomes that reflect meaningful dynamic aspects of economies. So rich in fact that here I could only present a few of the relevant results. I hope that these are convincing enough to show the potential of such a simulator to motivate game developers to work on them. Many aspects need further refinement of course. For example, the way I simulated credit does not reflect how credit normally works in an economy. However, the rudimentary credit simulated worked in synergy with several other aspects of the virtual economy.

A general insight from this exercise is that even in this simplest of meaningful economic model, important nonlinear instabilities were present, producing in many cases a non-intuitive dynamic outcome. This contrast with the widespread use of multiple linear regressions in classical economic analysis. If strong non-linear dynamics can be evidenced in very simple economic scenarios, then non-linear dynamics should be expected in real complex economic scenarios. Many examples of analysis of empirical data reveal this fact (see Jaffe et al 2013 for example).

A most remarkable aspect of Sociodynamica is that it captures the mysterious synergies achieved by the division of labor, as recognized long ago by Adam Smith. Modeling quantitatively the synergy that emerged when simulating division of labor could be very powerful in deepening our understanding of real economies. Few empirical research focuses on these aspects, as interest among classical economist and business managers seems to lay more on organization and management then on spontaneous emergence of order. Quantitative work on the effect of division of labor on the energy consumption of ants revealed its large potential economic impact (Jaffe & Heblin-Beraldo 1993). The fact that the insight by Adam Smith, that the "Invisible Hand" manages markets based on this division of labor, could be confirmed numerically with Sociodynamica is important. Division of labor unleashes synergies from the interactions of agents that cannot otherwise emerge. Focusing on the nature and dynamics of these synergies should be an important task for future economists. The present consensus is that strong institutions and rules favor the unleashing of synergies of a free market (World Bank 2010). Empirical evidence, however, shows that the relative importance assigned to basic scientific knowledge is a much better predictor of present economic strength and of future economic growth than any of the indexes estimating more classical economic variables like strength of institutions, industrial development, education and others (Jaffe et al 2013). That is, precision and depth in knowledge seems to as important or even more so than social and political structures in building economic synergies.

The emergence of collective phenomena have their own dynamics, as was illustrated here with the effect the asymmetry of price sensitivity between buyers and sellers has on the economic dynamics. Some examples from real life can illustrate this point. In most countries, drinkable water is more expensive than cheep wine or fizzy drinks, despite the fact that water is the more abundant resource. Results from Sociodynamica suggest that this phenomenon might be explained by putative asymmetries in price sensitivities between consumers need to quench their thirst and sellers managing their stock of water supply rationally. These asymmetries might explain the above average price inflation in health care where the urgency of buyers contrast with the secure position of sellers. The model shows the probable outcome of a combination of greedy health providers and frightened patients.

A relevant insight gained here relates potentially to insurance schemes. Sociodynamica showed that markets for risky utilities, such as provided by insurance, are nor efficiently tracked by simple market forces. This might explain the known differences in costs and efficiency of health insurance systems based on market forces and those based on central planning. Novel incentives to make insurance markets more rational might be tested and perfected with simulations before implementation.

These results just confirm Hayek's hunch that "the curious task of economics is to demonstrate to men how little they really know about what they imagine they can design." (Hayek 1991). Even very simple abstractions of economies produce emergent phenomena and irreversible dynamics that are more appropriately understood with scientific tools familiar to ecologist than to physicists. This would suggest that our future understanding of real economies will profit more from consilient bridges with the biological sciences, focusing on diversity and stability rather than econophysics or mathematics with its focus on precise quantitative predictions. This cocktail of approaches can be achieved with agent-based simulations as shown here. Our results show an example of the counterintuitive results of complex economic interactions. The differential effect of free markets on the average age of agents, and on the accumulated wealth of the society, mirrors the polemic between economist focusing on the collective benefits of policies (i.e. socialists for example) and those focusing on individual freedoms (liberal economist). Sociodynamica seems to be able to provide evidence of how both factors work simultaneously in a given society.

Eventually, it might be possible with more sophisticated models, to reflect real economic dynamics in a given system. Climate modeling and weather forecasting have taught us that the task of understanding reality can be tamed, partially with complex simulation modes.  Even if for specific settings and moments, Hayek's Economic Calculus might have a solution, the main lessons for economist is that economies are more like the weather, complex systems not easy to predict. Thus, interventions into the economic dynamic have to be treated as carefully experiments that have to be monitored and reversed if the outcome turns out to contradict expectations. Hopefully, letting central planners and interventionist prone politicians play games based on these simulations, will help them understand the nature of complex dynamics in economics.

# Appendix

**Sociodynamica** creates a virtual society where **a**gents exploit and compete for resources and share resource 1 among them, according to the settings defined by the internal parameters and the external parameters. The agents may acquire renewable and non-renewable resources trough work; they may accumulate those resources and commercialize them. At the same time, agent may acquire resources through commerce.

*Global Parameters*

**POP:**  Number of agents (no)

**TAR:**  Aggregate total wealth accumulated by all agents

*Simulation logic*

**Each time step:**
Do simulation loop
Matrix:
    Eliminate variable types previously defined
    Eliminate agents with wealth = 0 (par 11)
    Increment age agent(i, 7) = agent(i, 7) + 1
    Use of resource 1 and 2 agent(i,11)- BRC1; agent(i,12) – BRC2
    Assessment of GDP GDP = GDP + agent(i, 11)
    Show and Plot

*Plot* Shows the agents according to their total resources (Food+Money) indicated as the sqr of the diameter.
The high of the agent is proportional to the total wealth of the agent's money.
The color of the bubble depends of the type of agent as indicated at the left bottom of the screen
The thickness of the border is proportional to the perceived cost of living (Prize for food + Price for minerals).
The color of the border is more reddish or even yellow the higher the ratio MinPrice/FoodPrice. Blue borders indicates that food prices are higher than mineral prices.
Black bar at the bottom indicates a length of 100 pixel

**Internal parameters:**

*General parameters* (number in parenthesis indicates the column in the master matrix)

| | | |
|---|---|---|
| X (0) | Spatial dimension1 | |
| Y (1) | Spatial dimension2 | |
| CRa (2) | **Contact Radius or Contact** | Maximum distance at which interchange between agents m |

|             | **Horizon**            | example)                                                    |
| TMo (4)     | **Type of Movement**   | Type of spatial displacement: NO MOVEMENT                   |

*Characteristics of agents*

| Age (7)   | **Age**              | Age of agent                                                           |
| WT (10)   | **Wealth-Money**     | Total capital in liquid money                                          |
| WFo (11)  | **Wealth-Food**      | Amount of resource 1 (Renewable). If WFo(i) = 0 then age               |
| WCo(12)   | **Wealth-Commodity** | Amount of resource 2 (Minerals or Non-Renewable resour                 |
| Dept (13) | **Dept**             | Accumulated Dept                                                       |
| Well (19) | **Well-being**       | Amount of resources 3 = r1 * r2 * Gain                                 |

TAg (20)   **Type of Agent**   Specialization or task of agent
  0 Omnipotent: exploits resources 1 and 2
  1 Farmer: exploits only resources 1
  2 Miners: exploits only resources 2
  3 Trader: does not collect but trades and provides credit: i.e creates money

**External Parameters**

*General*

**Initial Nr. of Agents** (ino)

**Optimum Population Size** (ops): Maximum number of agents aimed at through ssconst

**Simulation Scenario**: production of new agents (ssconst):
   0: New agents are created each time steps, until the number indicated by ops is achieved.
   New agents are assigned internal parameters at random

**Proportion Culled** (PC). The proportion of agents killed randomly when population in excess of ops

**Dangers**, other than starvation danger Large values increase random selection; large WCo reduces this.

**Fitness function**: Agents, in order to continue in the virtual word, had to satisfy each time step the rule: 100 * Rnd / (amount of resource 2) / (mean wealth of resource 2) < 1000 * Rnd /

**Dangers:** probability of being eliminated in random selection events

*Resources 1 and 2*

**Resource 1 (provides WFo) and 2 (provides WCo)**
**Number of patches of Resource (**RNR) Number of resource patches
**Size of patch of Resource** (SNR) Maximum size of each patch (but see Mutation)
**Degradation of Resource** due to consumption (DNR) Amount lost due to consumption (RD)
**Distribution Pattern of Resource** (DPR) Resource is distributed:
  1: Fixed size, randomly distributed
  2: Fixed size, centered
  Else Random size, randomly distributed
**Basal rate of metabolism** (BRC) Amount of resource passively used-up (b)
**Efficiency of consumption** (EfC) Amount of resource assimilated.
   When EfC>10 then productivity is simulated: Agent collects = (EfC-10)*price (either for resource 1 and/or 2)
**Frequency of change in distribution (**FCh) Frequency in t-steps distribution changes
**Consumption of resource**: Rate of exploitation: resource = resource – DNR
   BRC: Wearing or passive use of resource agent- BRC
   Resource 1 can be modeled as a renewable resource (agriculture for example),
   whereas resource 2 as a non renewable resource (mining for example),
   by assigning DNR = 0 and DNR = 1 respectively

**Food Reserve** (FR): Minimum amount of food needed for agent to engage in transactions of any kind

**Min Food for Reproduction** (MFR): The amount of food that need's to be accumulated before reproduction can start when Simulation Scenario is 1 or 4

**Type of economy (EconoT)**
  **0**: No Barter nor any other interactions except taxes
  **1**: **Barter**: with no money
  **> 1**: Money as Species. If **Price adjust** = 0 then Fixed prices
  If **Price Adjust** > 0 the Prices are determined by demand: agents not selling decrease price by one unit; agents not finding seller from which to buy increase price by one unit.
  **3**: + Financial: Traders lend money.
  **4**: + Agents finishing a successful sale, increase price by one unit
  **5**: + Successful buyers decrease their future asking price by one unit
  **6**: + As in EconoT 4 and 5
  **7**: + As in 2, 3, 4 and 5
  8: As in 2 and 3
  **10:** Agents pay Taxes: Taxes collected are increased synergistically by Taxpool * SSTax prior to their distribution.

**Food price**: in integer units

**Mineral price**: Price of commodities in integer units

**Price Adjustment**: Units prices are reduced or increase when trade fails. 0 = fixed prices